\documentclass[aps,prd,reprint,showpacs,
groupedaddress,
nobibnotes,nofootinbib]{revtex4-1}
\usepackage{amsmath,amssymb,amsthm}
\usepackage{graphicx}
\usepackage{mathrsfs}
\usepackage{multirow}
\usepackage{hyperref}
\usepackage{color}

\newcommand{\beq}{\begin{equation}}
\newcommand{\eeq}{\end{equation}}
\newcommand{\bw}{\begin{widetext}}
\newcommand{\ew}{\end{widetext}}

\begin{document}

\let\thefootnote\relax\footnotetext{$^*$ Correspondence author: xxue@phy.ecnu.edu.cn}

\title{Effective Gravitational Theory at Large Scale with Lorentz Violation}

\author{Yiwei Wu$^{1}$,
Xun Xue$^{1 \, *}$,
Lixiang Yang$^{1}$, Tzu-Chiang Yuan$^{2}$}

\address{
$^{1}$Institute of Theoretical Physics, Department of Physics, East
China Normal University, No.500, Dongchuan Road, Shanghai 200241, China\\
$^{2}$Institute of Physics, Academia Sinica, Nangang, Taipei 11529, Taiwan
}

 %%%%%%%%%%%%%%%%%%%%%%%%%%%%%%%%%%%%%%%%%%%%%%%%%%%%%%%

\begin{abstract}

The dipole anomaly in the power spectrum of CMB may suggest that
the Lorentz boost invariance may be violated at cosmic scale.
Lorentz symmetry may not be an exact symmetry in Nature,
it may be partially broken at the galaxy scale.
We employ the symmetry of very special relativity as an example to
illustrate the Lorentz violation effects by constructing the corresponding
gauge theories as effective gravitational theories at large scale.
One common feature of this class of gravitation models is the
non-triviality of spacetime torsion and contorsion even if the matter
source is consisted solely of scalar fields. The presence of non-trivial contorsion
contributes to the effective energy-momentum distribution which may account
partly if not all for the dark matter in the universe.

\end{abstract}


\maketitle

{\it \underline{Introduction \& Motivation}} Lorentz symmetry is one of the most established exact symmetries of Nature,
withstand challenges by numerous experiments since the discovery of
special relativity by Einstein over one hundred years ago. However, the idea of minuscle
Lorentz violation attracts many attentions since 1990s. Possible Lorentz
violation (LV) effects might come from quantum gravity which are nevertheless
suppressed by Planck scale. On the other hand, all experimental results give very low
upper bound on possible Lorentz violation. At the macroscopic scale,
local Lorentz invariance is verified to very high accuracy within
solar system. On the other hand, at the cosmic scale, the dipole anomaly
in the power spectrum of cosmic microwave background (CMB) \cite{Planck2015,Kogut}
indicates that there exists a rest frame of CMB which breaks Lorentz
invariance even in empty space far away from any galaxy. There is
no direct signal on how the Lorentz symmetry behaves at large scale,
{\it e.g.} scale of or larger than the galaxy. Neither is there any experimental
test on it at such scale. On the contrary, there are some substantial
evidences in astronomical observations indicating implicit deviations
from predictions by general relativity or Newtonian gravitation theory,
such as the galaxy rotation curve and the accelerating expansion of
the universe, {\it etc.}

CMB also exhibits various anomalies at low multipoles and other puzzles,
such as very large scale anisotropies, anomalous alignments, and non-Gaussian
distributions \cite{Planck2015}. The quadrupole $(l = 2$, spherical
harmonic mode) has a low amplitude compared to the prediction of the standard
cosmological ($\Lambda$CDM) theory. In particular, the quadrupole and octupole ($l = 3$) modes appear
to have an unexplained alignment with each other and with both the
ecliptic and equinoxes -- this alignment was sometimes coined as
the {\it axis of evil}. It seems that there is a conflict between the cosmological
principle which states that the universe should be spatially isotropic and uniform,
and the principle of relativity at cosmic scale.
We are therefore facing the possibility of modifying one or even both of these principles
when encountering problems at cosmic scale.
Recall that when Einstein established his general relativity a century ago,
he was strongly motivated by Mach's principle. One of the very general statement of Mach's
principle is \textquotedblleft Local physical laws are determined
by the large-scale structure of the universe\textquotedblright. It
is therefore very inspiring to think about Mach's principle in formulating
an effective gravitational theory at large scale. Local Lorentz symmetry could be
violated at large scale due to {\it classical} gravitation effects.

We assume here that local Lorentz invariance is partly violated or partially broken
from the scale of galaxy to the cosmic scale, while it is exact within the
scale of solar system. In particular, Lorentz boost is violated in the cosmic
scale. An effective gravitation theory at large scale should be formulated
according to the above reasonings. How large of a region in spacetime
can be regarded as local when we talk about local symmetry?
It should be noted that the local requirements can actually be different depending on
different physical phenomena.
For the large scale gravitation effects, local may be very large compared
with the typical laboratory scale of Lorentz invariant electromagnetic processes.

There appear many attempts on investigation of possible Lorentz
violation from various theoretical aspects since the middle of 1990s. Coleman
and Glashow \cite{ColemanGlashaw} developed a perturbative framework to investigate the
deviation from Lorentz invariance in which the departure of Lorentz
invariance is parametrized in terms of a fixed timelike four-vector
or spurion. Colladay and Kostelecky \cite{Colladay} proposed
an extension of the standard model Lagrangian by incorporating
Lorentz and $CPT$ violation perturbations of more general spurion-mediated terms
induced by expectation values of Lorentz tensors due to spontaneous Lorentz symmetry breaking.
More recently,  Cohen and Glashow~\cite{CohenGlashow}, proposed the
very special relativity (VSR) in which
$CP$ violation is connected with Lorentz violation.
They argued that the local symmetry of known observations may not be
necessary as large as Lorentz symmetry but instead its subgroup would be enough
provided that the subgroup can be enlarged to the full Lorentz group when
discrete symmetry $P$, $T$ or $CP$ is incorporated. When $CP$ is a good symmetry,
the full spacetime symmetry is the complete Lorentz group. However,
if $CP$ is violated, as was observed in Nature, the spacetime symmetry may not be the complete Lorentz
group anymore and hence Lorentz violation may be connected with $CP$
violation. There are three subgroups of the Lorentz group that can be the symmetry
group of VSR. Denote the Lorentz group generators as $J_{x}$, $J_{y}$
and $J_{z}$, the rotation generators, and $K_{x}$, $K_{y}$ and
$K_{z}$, the Lorentz boost generators.
Up to isomorphism the three VSR symmetry groups are $E(2)$ and $Hom(2)$ with three generators,
and $Sim(2)$ with four generators. $E(2)$'s generators
are $T_{1}$, $T_{2}$ and $J_{z}$, $Hom(2)$'s generators are $T_{1}$,
$T_{2}$ and $K_{z}$, while $Sim(2)$'s generators are $T_{1}$, $T_{2}$,
$K_{z}$ and $J_{z}$.
Here $T_{1}=K_{x}+J_{y}$ and $T_{2}=K_{y}-J_{x}$ are the
two generators of an Abelian subgroup $T(2)$.

{\it \underline{Effective Gravity at Large Scale}}
To build the effective gravitational theory at large scale we employ
the VSR model as an example to introduce the Lorentz violation by
localizing the VSR symmetry to construct the VSR gauge theory as the
corresponding gravitation theory. The construction of gravitation theory
with gauge principle began just after the proposition of Yang-Mills
field theory in 1954. Utiyama  \cite{Utiyama} was the pioneer to extend
the gauge principle to non-compact Lie group, in particular the Lorentz group
and found that general relativity can be viewed as a Lorentz symmetry gauge theory.
Subsequently, more sophisticated Lorentz and Poincar$\rm {\acute e}$ gauge theories were formulated
by Sciama \cite{Sciama}, Kibble \cite{Kibble}, Ne$'$eman \cite{NeemanTrautman},
Trautman \cite{NeemanTrautman}, and Hehl \cite{F. W. Hehl} {\it etc}.
The bridge Einstein employed to make transition from special relativity to
general relativity was the equivalence principle, which
can be stated as one can always transform away the gravity effect
by choosing the appropriate reference frame at any point of spacetime.
The spacetime is locally flat in such a reference frame which
is known as the free falling reference frame. The orientation of coordinate
axes varies freely from one spacetime point to the other and has independent
locally Lorentz invariance. However, the locally flat coordinates
can not be holonomic globally for a general spacetime manifold but
are anholonomic coordinates in general. The coordinate transformation from locally
flat or free falling coordinates to a general holonomic coordinates
can be described by the tetrad or the vierbein fields $h_{a}^{\:\mu}$.
The relation between the tetrad fields and the metric tensor of a spacetime
manifold is
\begin{equation}
\eta_{ab}=g_{\mu\nu}h_{a}^{\:\mu}h_{b}^{\:\nu} \; ,
\end{equation}
and its inverse is
\begin{equation}
g_{\mu\nu}=\eta_{ab}h_{\:\mu}^{a}h_{\:\nu}^{b} \; .
\end{equation}
The commutator $\left[h_{a},h_{b}\right]=f_{ab}^{c}h_{c}$ for tetrad
basis $h_{a}=h_{a}^{\mu}\partial_{\mu}$ is non-trivial for anholonomic
coordinates in general.

At every spacetime point physics is dictated locally by special relativity
via a well-defined relativistic field theory Lagrangian in
terms of locally flat anholonomic coordinates. To transfer to the
case with gravity, the local Lorentz invariance is guaranteed by
introducing the Lorentzian gauge field, which behaves as the connection.
Suppose the matter field $\psi$ transforms according to the representation
of Lorentz group $U\left(\Lambda\left(x\right)\right)$ under the
local Lorentz transformation $x^{\mu}\rightarrow\Lambda_{\:\nu}^{\mu}\left(x\right)x^{\nu}$,
the local Lorentz invariance demands the Lagrangian in the Minkowski
spacetime background $\mathfrak{\mathcal{L}}\left(\partial_{\mu}\psi,\cdots\right)$
to transform into $\mathfrak{\mathcal{L}}\left(\mathcal{\mathfrak{\mathcal{D}}}_{\mu}\psi,\cdots\right)$,
one in the curved spacetime background, where the covariant derivative $\mathcal{D}_{\mu}$
is defined through the Lorentz Lie algebra $S_{ab}$ generators
as $\mathcal{D}_{\mu}=\partial_{\mu}-\dfrac{i}{2}{A^{ab}}_{\mu}S_{ab}$
by the gauge principle. Here ${{A}_{\mu }}=\dfrac{1}{2}{{A}^{ab}}_{\mu }{{S}_{ab}}$
is known as the Lorentzian gauge field or the Lorentzian connection.
If local Poincar$\rm{\acute e}$ invariance is taking into account, the tetrad fields
$h_{a}^{\:\mu}$ can be regarded as gauge potential of the local
translation of spacetime in some sense, while the curvature and torsion
are the field strengths for Lorentz connection and tetrad field respectively,
$\left[ {{D}_{a}},{{D}_{b}} \right]={{T}_{ab}}^{p}{{D}_{p}}+\dfrac{i}{2}{{R}_{ab}}^{pq}{{S}_{pq}}$.

The choice of gauge invariant action for the gauge field $A_{\,\,\,\mu}^{ab}$
and tetrad field $h_{a}^{\,\mu}$ in the locally Lorentz invariant
theory need to obey local Lorentz gauge transformation invariant and
local Lorentz invariant. Both Yang-Mills type and Hilbert-Einstein
one are acceptable. However, for action of Yang-Mills
type the coupling constant is dimensionless and there is no Newtonian gravity
limit, while for the action of Hilbert-Einstein type it can be proved to lead to general
relativity. The action for Poincar$\rm{\acute e}$ gauge theory can be also chosen
as Yang-Mills type
%which was discussed by Hehl  {\it et al.}
\cite{F. W. Hehl}, and for the
general linear group $GL(4,R)$ see \cite{Yang:2012pk}.
However, there are some ambiguities in the Yang-Mills type of non-compact gauge theory.

The starting point for the gravity part of Lorentz gauge theory is
then
\begin{equation}
S_{E}=\dfrac{1}{16\pi G}\int d^{4}xh{R_{ab}}^{ab} \;,
\end{equation}
where $h={\rm det}\left(h_{\,\,\mu}^{a}\right)$. The equations of motion (EoM)
for connection give a set of constraint equations for $A_{\,\,\,\mu}^{ab}$,
$\mathcal{D_{\nu}}\left(h\left({{h}_{a}}^{\nu }{{h}_{b}}^{\mu }-{{h}_{a}}^{\mu }{{h}_{b}}^{\nu } \right)\right)=0$,
which implies the Levi-Civita connection. The EoM for tetrad fields are
the Einstein field equations
\begin{equation}
{{R}^{a}}_{c}-\frac{1}{2}{{\delta }_{c}}^{a}R=0 \; .
\label{eq:einsteineq}
\end{equation}

In the presence of matter field, the theory is not torsion free in
general, the scalar source implies torsion free theory while spinor
and vector source gives non-trivial torsion distribution. However,
macroscopic sources are either scalar or non-polarized electromagnetic
field which lead to torsion free gravity, the general relativity.

To introduce Lorentz violation into the large scale effective gravitation
theory, we take $Sim\left(2\right)$ symmetry as an illustration.
For a local $Sim\left(2\right)$ theory, one just need to restrict
the local symmetry transformation on $Sim\left(2\right)$,
\begin{equation}
\psi \xrightarrow{{{x}^{\mu }}\to {{\Lambda }^{\mu }}_{\nu }{{x}^{\nu }}}U\left( \Lambda \left( x \right) \right)\psi ,
\;\;\; \Lambda \left( x \right)\in Sim\left( 2 \right) \; .
\end{equation}
In the case of $SIM\left(2\right)$, the gauge potential takes the
form
\begin{eqnarray}
  {{A}_{\mu }}
 &=& \frac{1}{2}{{A}^{ab}}_{\mu }{{S}_{ab}} \; , \nonumber\\
   &=& \frac{1}{2}\left( {{A}^{10}}_{\mu }+{{A}^{31}}_{\mu } \right){{T}_{1}}+\frac{1}{2}\left( {{A}^{20}}_{\mu }-{{A}^{23}}_{\mu } \right){{T}_{2}} \nonumber\\
   &&  +\, {{A}^{30}}_{\mu }{{K}_{3}}+{{A}^{12}}_{\mu }{{J}_{3}} \nonumber\\
   && +\, \frac{1}{2}\left( {{A}^{20}}_{\mu }+{{A}^{23}}_{\mu } \right)\left( {{S}_{20}}+{{S}_{23}} \right) \nonumber\\
    && + \, \frac{1}{2}\left( {{A}^{10}}_{\mu }-{{A}^{31}}_{\mu } \right)\left( {{S}_{10}}-{{S}_{31}} \right) \; .
\end{eqnarray}
The restriction on $Sim\left(2\right)$ implies
\begin{equation}
{{A}^{10}}_{\mu }-{{A}^{31}}_{\mu }=0 \, ,\ \ {{A}^{20}}_{\mu }+{{A}^{23}}_{\mu }=0 \; ,
\label{eq:constrain0}
\end{equation}
which can be realized by taking the action in the form
\begin{eqnarray}
S_{E} &=& \frac{1}{16\pi G}\int d^{4}xh\left({R^{ab}}_{ab} \right. \nonumber\\
 && +\left.\lambda_{1}^{\mu}\left({{A}^{10}}_{\mu }-{{A}^{31}}_{\mu }\right)+\lambda_{2}^{\mu}\left({{A}^{20}}_{\mu }+{{A}^{23}}_{\mu }\right)\right) \; ,
\end{eqnarray}
where $\lambda_{1}^{\,\,\mu}$ and $\lambda_{2}^{\,\,\mu}$ are the
Lagrange multipliers.
The EoM for connection give constraints on tetrad and contorsion
\begin{eqnarray}
  && \! \! \! \! D_{\nu}\left(h\left({{h}_{a}}^{\nu }{{h}_{b}}^{\mu }-{{h}_{a}}^{\mu }{{h}_{b}}^{\nu } \right)\right) \nonumber\\
   &&=\lambda_{1}^{\mu}h\left({{\delta }_{a}}^{1}{{\delta }_{b}}^{0}-{{\delta }_{a}}^{3}{{\delta }_{b}}^{1}\right)+\lambda_{2}^{\mu}h\left({{\delta }_{a}}^{2}{{\delta }_{b}}^{0}-{{\delta }_{a}}^{2}{{\delta }_{b}}^{3} \right) \; ,
\end{eqnarray}
which can be reduced to
\begin{equation}
\mathcal{D}_{\nu}\left(h\left({{h}_{1}}^{\nu }{{h}_{0}}^{\mu }-{{h}_{1}}^{\mu }{{h}_{0}}^{\nu } \right)\right)=-\mathcal{D}_{\nu}\left(h\left({{h}_{3}}^{\nu }{{h}_{1}}^{\mu }-{{h}_{3}}^{\mu }{{h}_{1}}^{\nu }\right)\right),\label{eq:constrain1}
\end{equation}
\begin{equation}
\mathcal{D}_{\nu}\left(h\left({{h}_{2}}^{\nu }{{h}_{0}}^{\mu }-{{h}_{2}}^{\mu }{{h}_{0}}^{\nu }\right)\right)=\mathcal{D}_{\nu}\left(h\left({{h}_{2}}^{\nu }{{h}_{3}}^{\mu }-{{h}_{2}}^{\mu }{{h}_{3}}^{\nu } \right)\right),\label{eq:constrain2}
\end{equation}
\begin{equation}
{\mathcal{D}_{\nu }}\left( h\left( {{h}_{\text{3}}}^{\nu }{{h}_{\text{0}}}^{\mu }-{{h}_{\text{3}}}^{\mu }{{h}_{\text{0}}}^{\nu } \right) \right)=\text{0}
\; ,
\label{eq:constrain3}
\end{equation}
and
\begin{equation}
{\mathcal{D}_{\nu }}\left( h\left( {{h}_{1}}^{\nu }{{h}_{2}}^{\mu }-{{h}_{1}}^{\mu }{{h}_{2}}^{\nu } \right) \right)=\text{0}
\; .
\label{eq:constrain4}
\end{equation}

The spin connection can be decomposed into torsionless part and contorsion
${{A}^{a}}_{bc}={{\tilde{A}}^{a}}_{\; \;bc}+{{K}^{a}}_{bc}$, where
$\widetilde{A}_{\; \;bc}^{a}=\dfrac{1}{2}\left( {{f}_{b}}{{^{a}}_{c}}+{{f}_{c}}{{^{a}}_{b}}-{{f}^{a}}_{bc} \right)$,
${{K}^{a}}_{bc}=\frac{1}{2}\left( {{T}_{b}}{{^{a}}_{c}}+{{T}_{c}}{{^{a}}_{b}}-{{T}^{a}}_{bc} \right)$
and ${T^{a}}_{bc}$ is the torsion tensor. The constraint relations
\eqref{eq:constrain1}, \eqref{eq:constrain2}, \eqref{eq:constrain3} and
\eqref{eq:constrain4} can reduce the number of independent components
of contorsion from 24 to 8, namely ${{K}^{10}}_{0},\ {{K}^{10}}_{1},\ {{K}^{10}}_{2},\ {{K}^{20}}_{0},\ {{K}^{30}}_{0},\ {{K}^{30}}_{1},\ {{K}^{30}}_{2},$ and ${{K}^{12}}_{0}$.
The constraint \eqref{eq:constrain0} reduces to
\begin{equation}
\begin{array}{l}
2{{f}^{\text{0}}}_{10}+{{f}^{0}}_{31}-{{f}^{1}}_{30}+{{f}^{\text{3}}}_{10}=0  \; , \\
  {{f}^{\text{1}}}_{10}+{{f}^{1}}_{31}=0  \; , \\
 {{f}^{\text{0}}}_{12}-{{f}^{\text{1}}}_{20}+{{f}^{\text{3}}}_{12}-{{f}^{2}}_{10}+{{f}^{1}}_{23}-{{f}^{2}}_{31}=0  \; , \\
  {{f}^{\text{1}}}_{30}+{{f}^{\text{0}}}_{31}+{{f}^{3}}_{10}+2{{f}^{3}}_{31}=0  \; , \\
  2{{f}^{\text{0}}}_{20}-{{f}^{\text{2}}}_{30}+{{f}^{3}}_{20}-{{f}^{0}}_{23}=0  \; , \\
  {{f}^{\text{2}}}_{10}+{{f}^{\text{2}}}_{31}+{{f}^{\text{0}}}_{12}+{{f}^{3}}_{12}+{{f}^{1}}_{20}-{{f}^{1}}_{23}=0  \; , \\
  {{f}^{2}}_{23}-{{f}^{2}}_{20}=0 \\
 {{f}^{\text{2}}}_{0\text{3}}+{{f}^{\text{0}}}_{2\text{3}}+\text{2}{{f}^{\text{3}}}_{2\text{3}}-{{f}^{\text{3}}}_{20}=0 \; .
\end{array}
\label{eq:sim2eq}
\end{equation}

The tetrad EoM takes the form of \eqref{eq:einsteineq}. For only the
symmetric part of connection can affect motion of particle through
the geodesic equation, we decompose the curvature
with the help of decomposition of spin connection as
\begin{equation}
{{R}^{mn}}_{ab}={{\tilde{R}^{mn}}}_{\;\;\;\;\;\;ab}+{{R}_{K}}{{^{mn}}_{ab}}+{{R}_{CK}}{{^{mn}}_{ab}} \; ,
\end{equation}
where ${{\tilde{R}^{mn}}}_{\;\;\;\;\;\;ab}$ and ${{R}_{K}}{{^{mn}}_{ab}}$ are
the curvatures composed of torsion-free connection and contorsion respectively,
while ${{R}_{CK}}{{^{mn}}_{ab}}$ contains cross terms of them. We can rewrite
\eqref{eq:einsteineq} as
\begin{equation}\label{EoMsim2}
 {{\tilde{R}}_{c}}^{a}-\frac{1}{2}{{\delta }_{c}}^{a}\tilde{R}=8\pi G{{\left( {{T}_{Sim\left( 2 \right)}} \right)}_{c}}^{a}
 \; ,
\end{equation}
where
\begin{eqnarray}
  {{\left( {{T}_{Sim\left( 2 \right)}} \right)}_{c}}^{a}  &=& \dfrac{1}{8\pi G}\left( \dfrac{1}{2}{{\delta }_{c}}^{a}\left( {{R}_{K}}+{{R}_{CK}} \right)\right. \nonumber\\
  &&-\left.\left( {{R}_{K}}{{_{c}}^{a}}+{{R}_{CK}}{{_{c}}^{a}} \right) \right) \; .
  \label{T-Sim2}
\end{eqnarray}

The force exerting on a particle moving in the gravitation field is supplied
by the torsion free part of the connection, the curvature of which satisfies
the Einstein field equation with the effective energy-momentum tensor
$T_{Sim\left(2\right)}$ of Eq.~(\ref{T-Sim2}) generated by an effective matter distribution
formally. It is worthy to note that $T_{Sim\left(2\right)}$ will
disappear if there is no matter distribution all over the space, {\it i.e.}
Minkowski spacetime is still the vacuum solution. To include the source
matter distribution, suppose the variation of the matter field action
is
\[
\delta {{S}_{M}}=\int{{{d}^{4}}x}h\left( \frac{1}{2}\delta {{h}^{c}}_{a}{{\left( {{T}_{M}} \right)}_{c}}^{a}+\delta {{A}^{ab}}_{\mu }{{\left( {{C}_{M}} \right)}_{ab}}^{\mu } \right) \; ,
\]
the complete gravitation equation will be
\begin{equation}\label{EoMcpl}
{{\tilde{R}}_{c}}^{\;\;a}-\frac{1}{2}{{\delta }_{c}}^{a}\tilde{R}=8\pi G{{\left( {{T}_{Sim\left( 2 \right)}}+{{T}_{M}} \right)}_{c}}^{a} \; .
\end{equation}

The effective energy-momentum tensor $T_{Sim\left(2\right)}$ contributes
to the gravitation in addition to matter contribution $T_{M}$
and appears as the dark side of the matter. Different source matter
distribution is expected to give rise different dark distribution.
The spherical or axial symmetric galaxy like source matter is expected
to lead to the possible contribution to dark matter effectively.

The complete constraints on the connection originated from \eqref{eq:constrain1},
\eqref{eq:constrain2}, \eqref{eq:constrain3} and \eqref{eq:constrain4}
become
\begin{eqnarray}
 {{\mathcal{D}}_{\nu }}\left( h\left( {{h}_{1}}^{\nu }{{h}_{0}}^{\mu }-{{h}_{1}}^{\mu }{{h}_{0}}^{\nu } \right) \right)+{{D}_{\nu }}\left( h\left( {{h}_{3}}^{\nu }{{h}_{1}}^{\mu }-\right.\right.&&\left.\left.{{h}_{3}}^{\mu }{{h}_{1}}^{\nu } \right) \right)   \nonumber\\
 =16\pi G\left[{{\left(C_{M}\right)}_{10}}^{\mu}+{{\left(C_{M}\right)}_{31}}^{\mu}\right] \; , &  & \label{eq:constrain1-1}
\end{eqnarray}
\begin{eqnarray}
  \mathcal{D}_{\nu}\left(h\left({{h}_{2}}^{\nu }{{h}_{0}}^{\mu }-{{h}_{2}}^{\mu }{{h}_{0}}^{\nu } \right)\right)-D_{\nu}\left(h\left( {{h}_{2}}^{\nu }{{h}_{3}}^{\mu }-\right.\right.&&\left.\left.{{h}_{2}}^{\mu }{{h}_{3}}^{\nu } \right)\right)   \nonumber\\
 =16\pi G\left[{{\left(C_{M}\right)}_{20}}^{\mu}-{{\left(C_{M}\right)}_{23}}^{\mu}\right] \; , && \label{eq:constrain2-1}
\end{eqnarray}
\begin{equation}
{{\mathcal{D}}_{\nu }}\left( h\left( {{h}_{3}}^{\nu }{{h}_{0}}^{\mu }-{{h}_{3}}^{\mu }{{h}_{0}}^{\nu } \right) \right)=16\pi G{{\left( {{C}_{M}} \right)}_{30}}^{\mu } \; , \label{eq:constrain3-1}
\end{equation}
and
\begin{equation}
 {{\mathcal{D}}_{\nu }}\left( h\left( {{h}_{1}}^{\nu }{{h}_{2}}^{\mu }-{{h}_{1}}^{\mu }{{h}_{2}}^{\nu } \right) \right)=16\pi G{{\left( {{C}_{M}} \right)}_{12}}^{\mu } \; ,\label{eq:constrain4-1}
\end{equation}
which lead to non-trivial torsion even in the scalar source-free case of
$C_{M}=0$ and hence the non-trivial effective energy-momentum
tensor $T_{Sim\left(2\right)}$ can contribute at least partly to possible dark matter
effect.

The discussions for other VSR symmetry groups are similar.
The constraint conditions for local $Hom(2)$ symmetry gauge theories are
\begin{equation}\label{hom2constrain}
{{A}^{10}}_{c}-{{A}^{31}}_{c}=0,\ \ {{A}^{20}}_{c}+{{A}^{23}}_{c}=0,\ \ {{A}^{12}}_{c}=0 \; ,
\end{equation}
while the 24 components of contorsion can be reduced to 12 independent ones, {\it e.g.}
\begin{eqnarray}
  {{K}^{10}}_{0},{{K}^{10}}_{1},{{K}^{10}}_{2}, && {{K}^{20}}_{0},{{K}^{20}}_{1}
,{{K}^{30}}_{0} \; , \nonumber\\
 {{K}^{30}}_{1},{{K}^{30}}_{2},{{K}^{12}}_{0},&& {{K}^{23}}_{1},
{{K}^{23}}_{3},{{K}^{31}}_{1} \; .
\end{eqnarray}
In the case of $C_M =0$, the 12 constraint conditions \eqref{hom2constrain} can be reduced to
\begin{equation}
\begin{array}{l}
{{K}^{10}}_{0}-{{K}^{31}}_{3}=2{{f}^{\text{0}}}_{10}+{{f}^{0}}_{31}-{{f}^{1}}_{30}+{{f}^{\text{3}}}_{10}  \; , \\
  {{K}^{20}}_{1}+{{K}^{23}}_{1}=\frac{1}{2}\left( {{f}^{\text{1}}}_{\text{20}}-{{f}^{\text{0}}}_{12}\text{+}{{f}^{2}}_{10}-{{f}^{1}}_{\text{23}}
  -{{f}^{\text{3}}}_{12}+{{f}^{2}}_{31} \right)\! \!, \\
  {{K}^{20}}_{0}+{{K}^{23}}_{3}=2{{f}^{\text{0}}}_{20}-{{f}^{0}}_{23}-{{f}^{\text{2}}}_{30}+{{f}^{3}}_{20}  \; , \\
  {{K}^{12}}_{0}=\frac{1}{2}\left( -{{f}^{0}}_{12}-{{f}^{\text{1}}}_{20}+{{f}^{\text{2}}}_{10} \right)  \; , \\
  {{K}^{30}}_{2}=\frac{1}{2}\left( 2{{f}^{\text{0}}}_{20}-{{f}^{0}}_{23}-{{f}^{\text{2}}}_{30}+{{f}^{3}}_{20}-2{{f}^{\text{1}}}_{12} \right)  \; , \\
  {{K}^{30}}_{1}=\frac{1}{2}\left( 2{{f}^{\text{0}}}_{10}+{{f}^{0}}_{31}-{{f}^{1}}_{30}+{{f}^{\text{3}}}_{10}+2{{f}^{\text{2}}}_{12} \right) \; ,
\end{array}
\label{eq:hom2eq1}
\end{equation}
and
\begin{equation}
\begin{array}{l}
{{f}^{\text{0}}}_{10}+{{f}^{0}}_{31}+{{f}^{\text{3}}}_{10}+{{f}^{3}}_{31}=0  \; , \\
  {{f}^{\text{1}}}_{10}-{{f}^{1}}_{31}=0  \; , \\
  {{f}^{1}}_{20}-{{f}^{1}}_{23}+{{f}^{\text{2}}}_{10}+{{f}^{2}}_{31}=0  \; , \\
  {{f}^{\text{0}}}_{20}-{{f}^{\text{0}}}_{2\text{3}}+{{f}^{\text{3}}}_{20}-{{f}^{\text{3}}}_{2\text{3}}=0  \; , \\
  {{f}^{2}}_{23}-{{f}^{2}}_{20}=0  \; , \\
  3{{f}^{1}}_{23}-2{{f}^{\text{1}}}_{20}+{{f}^{\text{2}}}_{31}-{{f}^{\text{3}}}_{12}=0  \; .
\end{array}
\label{eq:hom2eq2}
\end{equation}

For the $E(2)$ case, the constraint conditions \eqref{hom2constrain} become
\begin{equation}\label{e2constrain}
{{A}^{10}}_{c}-{{A}^{31}}_{c}=0 \, ,\ \ {{A}^{20}}_{c}+{{A}^{23}}_{c}=0 \, ,\ \ {{A}^{30}}_{c}=0 \, ,
\end{equation}
 and 12 independent components of contorsion can be chosen as
\begin{eqnarray}
  {{K}^{10}}_{0},\ {{K}^{10}}_{1},\ {{K}^{10}}_{2},&&{{K}^{10}}_{3},{{K}^{20}}_{0},\ {{K}^{20}}_{2} \;  , \nonumber\\
 {{K}^{20}}_{3},{{K}^{30}}_{0}, {{K}^{30}}_{1},&& {{K}^{30}}_{2},{{K}^{12}}_{0},\ {{K}^{31}}_{1} \; .
\end{eqnarray}
In the case of $C_M =0$, the 12 constraint conditions \eqref{e2constrain} can be reduced to
\begin{equation}
\begin{array}{l}
{{K}^{10}}_{0}-{{K}^{10}}_{3}=\frac{1}{2}\left( -2{{f}^{\text{0}}}_{10}-{{f}^{0}}_{31}+{{f}^{1}}_{30}-{{f}^{\text{3}}}_{10} \right)  \; , \\
 {{K}^{20}}_{3}-{{K}^{20}}_{0}=\frac{1}{2}\left( 2{{f}^{\text{0}}}_{20}-{{f}^{0}}_{23}-{{f}^{\text{2}}}_{30}+{{f}^{3}}_{20} \right) \; ,  \\
  {{K}^{10}}_{1}-{{K}^{31}}_{1}={{f}^{2}}_{23}-{{f}^{2}}_{20}  \; , \\
  {{K}^{30}}_{0}={{f}^{\text{0}}}_{30}  \; , \\
  {{K}^{30}}_{1}=\frac{1}{2}\left( {{f}^{\text{0}}}_{31}-{{f}^{1}}_{30}-{{f}^{\text{3}}}_{10} \right) \; ,  \\
  {{K}^{30}}_{2}=\frac{1}{2}\left( -{{f}^{\text{3}}}_{20}-{{f}^{0}}_{23}-{{f}^{\text{2}}}_{30} \right) \; ,
\end{array}
\label{eq:e2eq1}
\end{equation}
and
\begin{equation}
\begin{array}{l}
{{f}^{\text{0}}}_{10}+{{f}^{0}}_{31}+{{f}^{3}}_{10}+{{f}^{3}}_{31}=0 \; , \\
  {{f}^{\text{1}}}_{10}+{{f}^{1}}_{31}-{{f}^{2}}_{20}+{{f}^{2}}_{23}=0 \; , \\
  {{f}^{\text{1}}}_{\text{20}}-{{f}^{1}}_{\text{23}}\text{+}{{f}^{2}}_{10}+{{f}^{2}}_{31}=0  \; , \\
  {{f}^{\text{0}}}_{20}-{{f}^{0}}_{23}+{{f}^{3}}_{20}-{{f}^{\text{3}}}_{2\text{3}}=0  \; , \\
  {{f}^{\text{0}}}_{12}+{{f}^{\text{3}}}_{12}=0  \; , \\
  {{f}^{\text{0}}}_{30}+{{f}^{\text{1}}}_{10}+{{f}^{1}}_{31}+{{f}^{\text{3}}}_{30}=0 \; .
\end{array}\label{eq:e2eq2}
\end{equation}

We also take $T(2)$ symmetry into account, with 16 constraint
equations and 16 independent components of contorsion. We can arrive
at the conclusion that all VSR gauge theories are gravity theories with
non-trivial torsion in general. We propose that at least part of the dark matter effects
might be induced by contorsion contributions.

{\it \underline{Discussions}} In this paper, we begin with the analysis of dipole anomaly of CMB
power spectrum and speculate that at the cosmic scale boost invariance
is implicitly broken. However, at scale larger than the galaxy, the
boost invariance may not be broken totally. The VSR symmetry is illustrated
as an example of large scale local symmetry and its corresponding
gravity theory is constructed. Evidence of the Lorentz violation effect in all
these constructions is the existence of non-trivial torsion contributing
an effective energy-momentum distribution which may be regarded as the source from
the dark side of matter. The dark matter effect might be an emergent phenomenon
from large scale Lorentz violation in gravitation.
The implicit galaxy rotation curve based on the implicit solution
of contorsion in the presence of spherical or axial symmetric source
is still needed to be solved.

The approach discussed in this letter may also have cosmological application. There will
be an effective contorsion contribution, which may serve as the effective
dark energy, in the Robertson-Walker like solution based on cosmological
principle. Well known issues like the Boulware-Deser ghost \cite{BDghost} and the Zakharov discontinuity
\cite{Veltman, Zakharov} encountered in massive gravity theory are needed to be reexamined
in the present approach. Deser and Woodard \cite{deserwoodard}
proposed a non-local gravity model to take in account both dark matter as well as dark
energy. In our opinion, large scale behavior of gravity must have some
non-local features. It is also interesting to
investigate the relation between the two approaches.

At any rate, it will be very important to clarify the mechanism of Lorentz violation
at large scale. Does the Lorentz symmetry breakdown happen suddenly
at some scale, {\it e.g.} galaxy scale, or cascading or gradually depending
on some continuous parameters? How do we construct the model? Is the
origin of Lorentz violation at large scale the accumulation of other
shorter scales? Is there any relation between Lorentz
violating gravity and new physics beyond the standard model? We
would like to investigate and hopefully provide answers to some of these issues in the future.

{\it \underline{Acknowledgment}} We would like to thank
W. B. Yeung for helpful discussions. XX would like to thank the hospitality of the
Kavli Institute for Theoretical Physics China at the Chinese Academy of Sciences in Beijing
where this paper was completed.
This work is partially supported by the National Natural Science Foundation
of China, under Grant No. 11435005.

%%%%%%%%%%%%%%%%%%%%%%%%%%%%%%%%%%%%%%%%%%%%%

\bibliographystyle{unsrt}

\end{document}